\documentclass[preprint,showpacs,aip,pop,amsmath]{revtex4-1}
\usepackage{graphicx}
\usepackage{amsmath}
\usepackage{subfig}
\usepackage{bm}
\usepackage{verbatim}

\newcommand{\ExB}{$\bm{E}\times\bm{B} \,$}

\newcommand{\bhat}{\bm{\hat{b}}}

\newcommand{\xhat}{\bm{\hat{x}}}
\newcommand{\yhat}{\bm{\hat{y}}}
\newcommand{\zhat}{\bm{\hat{z}}}

\newcommand{\pfrac}[2]{\frac{\partial#1}{\partial#2}}

\newcommand{\curl}[1]{\nabla \times #1}
\newcommand{\np}{\nabla_{\perp}}

\newcommand{\reff}[1]{(\ref{#1})}

\renewcommand{\vec}[1]{{\mathbf{#1}}}

\newcommand{\dV}{d\bm{x}\,}

\begin{document}
\preprint{}

\title{Radial convection of finite ion temperature, high amplitude plasma blobs}
\author{M. Wiesenberger}
\email{Matthias.Wiesenberger@uibk.ac.at}
\affiliation{Institute for Ion Physics and Applied Physics, Association EURATOM-\"OAW,  University of
   Innsbruck, A-6020 Innsbruck, Austria}
\author{J. Madsen}
\affiliation{Association EURATOM-DTU, Technical University of Denmark, Department of Physics, 
2800 Kgs. Lyngby, Denmark}
\author{A. Kendl}
\affiliation{Institute for Ion Physics and Applied Physics, Association EURATOM-\"OAW,  University of
   Innsbruck, A-6020 Innsbruck, Austria}
 
\begin{abstract}
    We present results from simulations of seeded blob convection in the scrape-off-layer of magnetically confined fusion plasmas.  We consistently incorporate
    high fluctuation amplitude levels and finite Larmor radius (FLR) effects using a fully nonlinear global gyrofluid model. This is in line with conditions found in tokamak scrape-off-layers (SOL) regions.
    Varying the ion temperature, the initial blob width, and the initial amplitude, we found an FLR dominated regime 
    where the blob behavior is significantly different from what is predicted by cold-ion models. 
    The transition to this regime is very well described by the ratio of the ion gyroradius to the characteristic gradient scale length of the blob. 
    We compare the global gyrofluid model with a partly linearized local model. For low ion temperatures we find that simulations of the global model show more coherent blobs with an increased cross-field transport compared to blobs simulated with the local model. The maximal blob amplitude is significantly higher in the global simulations than in the local ones. When the ion temperature is comparable to the electron temperature, global blob simulations show a reduced blob coherence and a decreased cross-field transport in comparison with local blob simulations.
\end{abstract}

\maketitle

\section{Introduction} \label{sec:introduction}
Radially propagating filaments elongated along magnetic field lines are responsible for a major part of particle density, momentum, and energy cross-field transport in the scrape-off-layer (SOL) in Tokamaks\cite{Garcia2009, myra_rev_2011, IonitaNF2012}. These filaments are widely known as blobs in L-mode operation and ELM filaments in H-mode operation.
The particle density amplitude of such structures compared to the background density can be well above unity  \cite{ ENDLER1995, Zweben2002, Nold2010, IonitaNF2012, myra_rev_2011}. 
This can be seen as a consequence of the non-local nature
of blobs. Blobs are born in the vicinity of the last closed flux surface, where the plasma is denser, hotter, and has steeper gradients than in the SOL region\cite{Garcia2007, Xu2010}.
Furthermore, in the SOL region the ion temperature can be equal to or even higher than the 
electron temperature \cite{Adamek_2008, Uehara_1998, Reich2004, Kocan2012}.

Despite these facts, most existing simulations of seeded blob dynamics are based on models invoking a thin layer approximation \cite{ Krasheninnikov2001368, jmad2011FLRBlob, Garcia_Bian_Fundamensky_POP_2006, garcia_inter, garcia_blob}.  Essentially, the thin-layer approximation linearizes the charge balance equation assuming that the ion mass entering the polarization density is constant. 
Sometimes this approximation is called Boussinesq-approximation, a term more
commonly found in the context of thermal convection in ordinary fluids.
In fact, there are close similarities between thermal convection in fluids and the interchange motion
in magnetically confined plasmas\cite{garcia_inter}.
We refer to these models as ``local'' models.
The linearization is done to avoid severe costs in runtime and/or major challenges in algorithmic development for the solution of the nonlinear polarization equation in the 
form of a generalized Poisson problem. For this kind of problem fast fourier methods, 
which are highly effective for linear problems, are inefficient. 
Our work is based on a ``global'' model derived from the full-F gyrokinetic equations \cite{jmadGF2013} retaining the full nonlinear polarization density.
We use discontinuous Galerkin methods\cite{Cockburn1998, Yadav2013, Einkemmer2013} to discretize spatial derivatives.
These methods have been developed during the last decades and received increasing attention from the numerical community\cite{Cockburn2001}. They are very versatile in the choice of the desired order of accuracy, and they retain a high degree of parallelism in the resulting algorithm. 
We exploit this in an implementation for GPUs and are thus able to efficiently
solve the nonlinear polarization equation in each timestep. 

In the past mostly local drift-fluid models without FLR effects were used for seeded blob simulations \cite{Garcia_Bian_Fundamensky_POP_2006, Angus2012}.
Yet, there has also been efforts to incorporate the fully nonlinear
polarization density \cite{Yu2006, Angus2014}, or at least a reduced form of it\cite{Kube2011, Kube2012}, into these models. 
Ref. \cite{Yu2006, Angus2014} showed that the cross-field transport  
is enhanced by the 
nonlinear polarization equation compared to its reduced form. In $3$D simulations the blob
is affected by drift-waves, which dominate the cross field
transport \cite{Angus2012, Angus2014}. 
Ref. \cite{Kube2011, Kube2012} focussed on deriving scaling laws for the blob velocity,
which for small amplitudes increases with the square root of blob width and amplitude.
Moreover, the effects of sheath dissipation and dynamical friction on blob motion were investigated.
Ref. \cite{Manz2013} estimated the velocity scalings for warm ions. 
None of these works, however, discussed energetic consistency of the underlying model.

The influence of FLR effects on the convection of seeded blobs was investigated in Ref.\cite{jmad2011FLRBlob}. A local, energetically consistent gyrofluid model was used. It was shown that FLR effects can have a profound influence on the cross-field blob transport in certain parameter regimes. 
In particular, FLR effects brake the poloidal up-down symmetry 
in the particle density field and reduce fragmentation compared 
to the zero Larmor radius limit.

Here, we present seeded blob simulations using a global gyrofluid model including FLR effects, which allows studies of the cross-field transport of high amplitude, finite ion temperature blobs.
We investigate transport properties and, furthermore, compare our global model with a local model in order to 
test the validity of the thin-layer approximation.

This paper is organized as follows: In section \ref{sec:global} we introduce
the ``global'' gyrofluid model equations as well as a mass and an energy 
theorem.  We then discuss ``local'' model equations in section \ref{sec:local} that we use to investigate the implications of lifting the thin-layer approximation
 and derive
the correspondence to existing isothermal drift-fluid
models in \ref{sec:comparison}. In section \ref{sec:simulations} we present results
of seeded blob simulations. In section \ref{sec:zlr} we discuss the
cold ion limit, in which FLR effects are eliminated. Then we explore the parameter range  
where FLR effects dominate the blob evolution in section \ref{sec:flr}.  
We present results of global, hot ion, and high amplitude simulations in section \ref{sec:high}. 
We conclude in \ref{sec:conclusion}.

\section{Gyrofluid models}\label{sec:equations}
Gyrofluid models\cite{knorr_1988,dorland-hammet,scott_pop_2010,jmadGF2013} emerge when taking gyrofluid moments of the gyrokinetic Vlasov-Maxwell equations\cite{brizard_Hahm_review2007}. Gyrokinetic models describe low-frequency turbulence in strongly magnetized plasmas. Gyrokinetic theory was developed to decouple the fast gyration time-scale present in turbulent fusion plasmas while retaining important finite Larmor radius (FLR) effects and thereby significantly reduces the computational requirements for numerical simulations.
The exact gyrokinetic system is highly complex, so for practical applications\cite{Dubin1983} limiting forms are used. Generally, two paths have been pursued: 1) delta-F models, in which gyrokinetic distribution functions are split into stationary background and small perturbed parts and 2) full-F models, in which finite Larmor radius (FLR) corrections to the polarization and magnetization densities in Maxwell equations are neglected, but in which the gyrokinetic distribution functions are not linearized. No a priori assumptions about fluctuation amplitudes are made in full-F models. Full-F models are therefore well suited for studies of edge and scrape-off-layer turbulence and the associated transport in magnetically confined fusion plasmas. 

\subsection{Global gyrofluid model}\label{sec:global}
Here, we will use a gyrofluid model\cite{jmadGF2013} derived from the full-F gyrokinetic model. The gyrofluid model retains all relevant nonlinearities including the full nonlinear polarization density, while also retaining FLR  effects. The gyrofluid model therefore allows us to investigate the interchange dominated convection of plasma filaments having large amplitudes and finite ion temperatures. We restrict ourselves to a simple paradigmatic two-field model, which describes the time evolution of the electron particle density $n$ and the ion gyrocenter density $N$ in a simple, quasi-neutral, isothermal, electrostatic plasma in the plane perpendicular to the magnetic field $\bm{B}$ at the outboard midplane. Parallel dynamics along magnetic field lines as well as sheath boundary physics are absent from the model. We employ a right-handed slab geometry with orthonormal unit vectors $(\xhat,\yhat,\zhat)$ with $\zhat$ aligned with the magnetic field and $\xhat$ anti-parallel to the magnetic field gradient. The inverse magnetic field strength is given 
as $\frac{1}{B} = \frac{1}{B_0}\left(1 +  \frac{x}{R}\right)$, where $R$ is the radial distance to the inner edge of the plane at the outboard mid-plane. The equations appear as 
\begin{subequations}\label{eq:GF_slab} 
\begin{align}
      \pfrac{n}{t} 
      +\frac{1}{B}\{\phi,n\}
      + n \mathcal{K}(\phi) 
      - \frac{T_e}{e} \mathcal{K}(n)      
      &= \nu \nabla_{\perp}^2 n  ,\label{eq:Elec_Cont_slab}\\
    \pfrac{N}{t} 
      +\frac{1}{B}\{\psi,N\}
      + N \mathcal{K}(\psi) 
      + \frac{T_i}{e} \mathcal{K}(N)      
      &= \nu \nabla_{\perp}^2 N  ,\label{eq:Ion_Cont_slab}\\      
       \Gamma_1 N+\nabla\cdot \bigg( \frac{N}{\Omega B}\nabla_\perp \phi \bigg) = n,  \label{eq:QN_slab}
   \end{align}
\end{subequations}
where $T_e$ and $T_i$ denote electron and ion temperature, respectively, $\nu$ is the collisional diffusion coefficient, $\Omega = \frac{eB}{m_i}$, and $\nabla_{\perp} = -\zhat \times (\zhat \times\nabla)$.
The \ExB-advection terms are written in terms of Poisson brackets, which for two arbitrary functions $f$ and $g$ are defined as 
\begin{align}
    \{f, g\}   = \pfrac{f}{x}\pfrac{g}{y}-\pfrac{f}{y}\pfrac{g}{x}.
\end{align}
The compressibility of the perpendicular fluxes is described by the operator
\begin{align}
  \mathcal{K} = -\kappa\pfrac{}{y},
\end{align}
with $\kappa = 2/(B_0R)$.
The third and fourth terms on the left hand side of Eq. \reff{eq:Elec_Cont_slab} represent the compression of the \ExB and the electron grad-B particle-density-fluxes, respectively. The latter is equivalent to the compression of the electron diamagnetic particle density flux, which is only finite when the magnetic field is inhomogeneous. 

Ion FLR effects appear in the quasi-neutrality constraint Eq.~\reff{eq:QN_slab} and in the generalized ion \ExB-velocity explicitly through the Pad\'e approximant $\Gamma_1 = \big( 1-\frac{1}{2}\rho_i^2 \Delta \big)^{-1}$ to the gyroaveraging operator\cite{dorland-hammet}, where $\rho_i = \sqrt{\frac{T_i}{m_i\Omega_0^2}}$ denotes the thermal ion gyroradius with the constant ion gyrofrequency $\Omega_0 = eB_0/m_i$. The gyroaveraging operator $\Gamma_1$ enters the generalized ion \ExB-velocity through the generalized potential $\psi := \Gamma_1 \phi - \frac{m}{2q} | \bm{u}_E|^2$, where $\bm{u}_E = \frac{\zhat \times \nabla \phi}{B}$ denotes the \ExB-velocity. 
The second term on the left hand side of the quasi-neutrality constraint Eq.~\reff{eq:QN_slab} is the nonlinear polarization density, which is the gyrofluid representation of ion inertia i.e. the ion polarization drift. 
The first term is the gyroaveraged charge contribution of ion gyroorbits belonging to gyrocenters described by $N$. 
The right hand side describes the electron charge contribution.

The time-evolution of the total particle and ion gyrocenter densities is governed by
\begin{align}
    \frac{d}{dt}\int_D \dV n &= \nu\int_D\dV \nabla_\perp^2 n,\\
    \frac{d}{dt}\int_D \dV N &= \nu\int_D\dV \nabla_\perp^2 N,
    \label{eq:Mass_balance}
\end{align}
where $D$ is the total simulation domain.
In the absence of diffusion, $n$ as well as $N$ are therefore conserved. 

To derive the energy conserved by the gyrofluid equations \reff{eq:GF_slab}, the electron particle density equation \reff{eq:Elec_Cont_slab} is multiplied by $T_e(1+\ln n) - e\phi$ and is integrated over space. In the same way, the ion gyrocenter density equation \reff{eq:Ion_Cont_slab} is multiplied by $T_i(1+\ln N) + e\psi$ and is integrated over space. The equations are integrated by parts and surface terms are dropped. Note that the gyroaveraging operator $\Gamma_1$ is self-adjoint. Summing the resulting equations and using the quasi-neutrality constraint Eq.~\reff{eq:QN_slab}, the energy invariant becomes
\begin{align}
      \frac{d}{dt} \int_D d \bm{x}\, \left(U_e + U_i + U_E \right)
      = \int_D d\bm{x}\, U_{\Lambda}.
\label{eq:Energy_balance}
\end{align}
The electron Helmholtz free energy $U_e$ and the ion Helmholtz-free-like energy $U_i$ are given as
\begin{align}
  U_e =  T_e n \ln n, \quad  
  U_i =  T_i N \ln N.
\end{align}
The ion gyrocenter density $N$ can be expressed in terms of $n$ and $\phi$ through the quasi-neutrality constraint Eq.~\reff{eq:QN_slab}. Therefore, $U_i$ describes ion Helmholtz free energy only to lowest order and will inevitably also include $\phi$-dependent terms. The \ExB-energy is defined as 
\begin{align}
    U_E = \int_D d \bm{x}\, \frac{1}{2}m_i N u_E^2.
\end{align}
An essential observation is that the full ion gyrocenter density $N$ enters $U_E$. In delta-F based models the ion gyrocenter density entering the \ExB-energy is constant and hence weighs all ion gyrocenter densities equally. This approximation is crude in the presence of high amplitude plasma filaments. Finally, energy dissipation due to particle density diffusion and ion gyrocenter diffusion becomes
\begin{align}
    U_{\Lambda} =  \int_D d\bm{x}\,     
    \big[e\psi + T_i(1+\ln N)\big] \nu \nabla_{\perp}^2 N
    -\big[e\phi - T_e(1+\ln n)\big] \nu \nabla_{\perp}^2 n. 
\end{align}

\subsection{Local gyrofluid model}\label{sec:local}
In most previous works local models were used to investigate the convection of seeded blobs\cite{Garcia_Bian_Fundamensky_POP_2006,myra_rev_2011, jmad2011FLRBlob, garcia_inter}. Here, we denote a model ``local'' when the polarization density is linearized.  In order to quantify how the nonlinear polarization influences blob convection and in order to determine in which regimes local models are valid, we compare the global model Eqs.~\reff{eq:GF_slab} with the following local gyrofluid model\cite{jmad2011FLRBlob}
\begin{subequations}\label{eq:GF_local} 
\begin{align}
    \pfrac{ \tilde n }{t} + \frac{1}{B_0}\{\phi,\tilde n\} + n_{0} \mathcal{K}(\phi) - \frac{T_e}{e} \mathcal{K}(\tilde n) &= \nu \np^2 \tilde n,\label{eq:Elec_Cont_local}\\
    \pfrac{\tilde N}{t} + \frac{1}{B_0}\{\Gamma_1\phi,\tilde N\} + N_{0} \mathcal{K}(\Gamma_1\phi) + \frac{T_i}{e} \mathcal{K}(\tilde N) &= \nu \np^2 \tilde N,\label{eq:Ion_Cont_local}\\
     \Gamma_1 \tilde N 
     +\frac{eN_{0}}{T_i} (\Gamma_0 - 1)\phi = \tilde n \label{eq:QN_local},
\end{align} 
\end{subequations}
where the gyroaverage operator $\Gamma_0 = \big( 1-\rho_i^2 \np^2 \big)^{-1}$ describes local finite inertia effects as well as higher order FLR corrections to the polarization drift\cite{jmad2011FLRBlob}; $n_0$ = $N_{0}$ denote constant reference particle and ion gyrocenter densities, respectively. We explicitly denote the local electron and ion gyrocenter densities $\tilde n$ and $\tilde N$ in order to distinguish local and global gyrofluid models. We stress that the thin-layer approximation is invoked in the model, which can be seen from the polarization density in Eq.~\reff{eq:QN_local}, which in the long wavelength limit (LWL) equals $eN_{0}T_i^{-1} (\Gamma_0 - 1)\phi \simeq e N_{0}\np^2 \phi$. In the absence of collisional effects the local gyrofluid model\cite{scott:102318,jmad2011FLRBlob} is a superset of local drift fluid models e.g. \cite{garcia_inter,Krasheninnikov2001368}. More detailed comparisons between local and global gyrofluid models as well as drift fluid models will be given in the next sections.   

\subsection{Local and global models} \label{sec:comparison}
Gyrofluid models are remarkably simple compared with drift fluid models, which include FLR effects e.g.\cite{Smolyakov_1998, Chang_Callen}. The reason why gyrofluid models are able to retain relatively simple functional forms is that much of the complexities associated with FLR effects have been incorporated into the gyrofluid moments themselves through the underlying gyrocenter coordinate transformation. The downside to the simple functional forms is that the corresponding gyrofluid moments do not directly describe well-known physical quantities like particle density, electric potential etc. Consider the global quasi-neutrality constraint Eq.~\reff{eq:QN_slab}. It is clear that we cannot express $N$ in terms of $n$ and $\phi$ on a closed form. However, in the long wavelength limit (LWL) we obtain 
\begin{align}
    N = n - \frac{\rho_i^2}{2} \np^2 n - \nabla \cdot \bigg(\frac{n}{\Omega B} \np \phi \bigg),
    \label{eq:N_LWL}
\end{align}
demonstrating that $N$ depends on particle density, the magnetic field-aligned component of the \ExB-vorticity, and the ion diamagnetic vorticity\cite{jmad2011FLRBlob}. Therefore, it is important always to keep the composite nature of gyrofluid moments in mind whenever gyrofluid models are used to describe plasma dynamics and when gyrofluid models are compared with other models. 

To obtain a clearer picture of the dynamics described by the global gyrofluid model given in Eq.~\reff{eq:GF_slab}, we derive a charge continuity equation. The charge continuity equation describes the time-evolution of the magnetic field aligned component of the \ExB-vorticity $\zhat \cdot \curl \bm{u}_E$ and is therefore often referred to as the \textit{vorticity equation}. This global LWL vorticity equation is derived by taking the time derivative of the quasi-neutrality equation \reff{eq:QN_slab} using Eq.~\reff{eq:N_LWL} to eliminate $N$
\begin{align}
    \nabla \cdot \bigg( \frac{n}{\Omega B} \big[\pfrac{}{t} + \frac{1}{B}\{\phi,\}\big]\nabla_{\perp}\phi^*\bigg)
    = \frac{T_e + T_i}{e}\mathcal{K}(n).
    \label{eq:vorticity}
\end{align}
Here, diffusive terms are neglected and we have defined 
\begin{align}
    \phi^*
    = \phi 
      +
      \frac{T_i}{e}\ln n.
\end{align}
The vorticity equation shows that the global gyrofluid model is a superset of corresponding global drift fluid models\cite{Zeiler,Kube2011} in the absence of collisions. 

Similarly, for the local gyrofluid model Eqs.~\reff{eq:GF_local} the approximate LWL representation of the ion gyrocenter density becomes 
\begin{align}
    \tilde N = \tilde n - \frac{\rho_i^2}{2} \np^2 \tilde n - \frac{n_0}{\Omega_0 B_0}\np^2 \phi, 
    \label{eq:N_LWL_local}
\end{align}
which can be used to derive the local LWL vorticity equation 
\begin{align}
      \nabla \cdot \bigg(\frac{n_0}{\Omega_0 B_0 }\big[\pfrac{}{t} + \frac{1}{B_0}\{\phi,\}\big]\nabla_{\perp} \tilde{\phi}^*\bigg) = \frac{T_e + T_i}{e} \mathcal{K}(\tilde n),         
\label{eq:VortLWL_DF_local}
\end{align}
where 
\begin{align}
    \tilde{\phi}^* := \phi + \frac{T_i}{e}  \frac{\tilde n}{n_0} .
  \label{eq:gradPhiLocal}
\end{align}
The local vorticity equations equals the drift-fluid vorticity equation\cite{Hinton_Horton_1971,scott:102318,Belova_2001} in the absence of collisions, showing that the local gyrofluid model is a superset of corresponding local drift-fluid models.  

The right hand sides of the LWL local Eq.~\reff{eq:VortLWL_DF_local} and global Eq.~\reff{eq:vorticity} vorticity equations are identical. The right hands sides describe the compression of the  electron and ion diamagnetic fluxes and transfer energy between Helmholtz free energy end kinetic energy\cite{Scott_2005}. 

The left hand sides describe the compression of the ion polarization flux, which consists of the magnetic field aligned components of \ExB-vorticity and ion diamagnetic vorticity. The ion diamagnetic vorticity, i.e. the ion pressure dependent part, can be shown to be the manifestation of LWL FLR effects\cite{scott:102318,jmad2011FLRBlob} in the vorticity equations. In the local model Eq.~\reff{eq:VortLWL_DF_local} the particle density is taken as a constant. This has two immediate consequences. First, the nonlinearity $\propto \nabla n \cdot \nabla \phi$ entering the global vorticity equation is absent in the local model. The implications of this ``thin-layer'' approximation is a priori difficult to predict. In the local model, if the ions are cold, the early blob evolution is characterized by a poloidal dipole structure in the electric potential, which is $\pi/2$ phase shifted with respect to the density field. Therefore, one could expect that the nonlinearity in the initial phase plays a minor role. When the ion temperature is finite, the dipole part of the electric field is accompanied by an electric field, which circumferences the density field representing FLR effects\cite{jmad2011FLRBlob}. Therefore, the nonlinearity is expected to influence the blob convection even in the initial phase when the ion temperature is finite. 

Second, in the local model vorticity is everywhere weighted by $n_0$, which implies that plasma inertia is everywhere constant and therefore independent of the local plasma density. This approximation enters the ``inertial'' blob velocity scaling estimated by dimensional analysis\cite{garcia_inter, Kube2011,jmad2011FLRBlob}, which in previous works has shown good agreement with numerical simulations in the high Reynolds number regime. Neglecting the nonlinearity, the inertial scaling emerges by balancing the electric field dependent part of the vorticity with the compression of the diamagnetic flux. The resulting local and global perpendicular velocity scalings become
\begin{subequations}
\begin{align}
    V_{ \text{local}} &= c_s\sqrt{\frac{\sigma}{R}\frac{\Delta n}{n_0}},\label{eq:velocity_scaling_local}    \\
    V_{\text{global}} &= c_s \sqrt{\frac{\sigma}{R}\frac{\Delta n}{(n_0+\Delta n)}}. \label{eq:velocity_scaling_global}
\end{align}
\label{eq:velocity_scaling}
\end{subequations}
Here, $\Delta n$ is the blob amplitude, $c_s = \sqrt{n_0(T_e+T_i)/m_i}$ is the acoustic speed, and $\sigma$ denotes the characteristic blob size. 
Eq. \reff{eq:velocity_scaling} also defines the interchange rates
\begin{align}
    \gamma_\text{local} = \frac{V_{\text{local}}}{\sigma} \quad\text{and}\quad
    \gamma_\text{global} = \frac{V_{\text{global}}}{\sigma}.
    \label{eq:interchange_rates}
\end{align}
The global scaling reduces to the local velocity scaling\cite{Kube2011} for small perturbation amplitudes $\Delta n /n_0 \ll 1$, which predicts a scaling $V_{\perp}/c_s \propto \sqrt{\Delta n}$. The local and global scalings predict very different blob velocities when $\Delta n /n_0 \geq 1$. The local scaling does not differentiate small or high perturbation amplitudes, whereas the global scaling predicts that the blob velocity asymptotically approaches $c_s \sqrt{\sigma/R}$. 

Another difference between the global and the local models is that the diamagnetic part of the vorticity is linearized in the local model (see Eq.~\reff{eq:gradPhiLocal}), whereas the corresponding diamagnetic term in the global model has a logarithmic dependence. Since the diamagnetic vorticity is the representation of FLR effects in the vorticity equation, the local model could potentially overestimate the importance of FLR effects in the presence of high fluctuation amplitudes. 

Finally, we note a distinct difference between the local and the global model regarding
the extent to which FLR corrections are made to the polarization density. 
By taking the low-amplitude limit of the global polarization equation 
\reff{eq:QN_slab}, the local polarization equation \reff{eq:QN_local} is not recovered because FLR corrections residing in the ``$(\Gamma_0-1)$'' operator in the local quasi-neutrality constraint Eq.~\reff{eq:QN_local} are not included in the global model. The local model is therefore more precise than the global model when gradient length scales are comparable to the ion gyroradius and amplitudes are small. Gyrokinetic models, which can handle large fluctuations amplitudes and gradient length scales comparable to the ion gyroradius, have been formulated\cite{dimits:055901}. However, compared with traditional nonlinear gyrokinetic models, these extended models are significantly more complex. Gyrofluid models based on extended gyrokinetic models have not been derived yet.

\section{Simulations} \label{sec:simulations}
In this section we present results from numerical simulations of the local gyrofluid model Eqs.~\reff{eq:Elec_Cont_local}-\reff{eq:QN_local} and the global gyrofluid model Eqs.~\reff{eq:Elec_Cont_slab}-\reff{eq:QN_slab}.
All results in this section describe simulations of blobs initialized as
\begin{align}
    n( x,y,0) = \Gamma_1 N(x,y,0) = n_0 + \Delta n \exp\left(- \frac{(x-x_0)^2+(y-y_0)^2}{2\sigma^2} \right) ,
    \label{}
\end{align}
where $\sigma$ is the initial blob width, ($x_0$, $y_0$) the initial position, and $\Delta n$ the initial blob amplitude. 
In this way the potential $\phi(x,y,0) = 0$ via the polarization equation. 
The simulation domain is a square box $D := [0,L] \times [0,L]$, where the box size is set to  $L=40\sigma$ in order to mitigate the influence of the boundaries. For the global gyrofluid model the $y$ boundaries are periodic, whereas Dirichlet boundary conditions are chosen at the $x$ boundaries
\begin{subequations}
\begin{align}
    n( 0,y,t) = n( L, y,t) = N( 0,y,t) = N( L, y,t) = n_0, \\
    \phi( 0,y,t) = \phi( L, y,t) = 0.
\end{align}
\end{subequations}
The local gyrofluid model is solved on a doubly periodic domain. 
 
In order to solve Eqs. \reff{eq:GF_slab}, we use discontinuous Galerkin (dG) methods
\cite{Cockburn2001, Cockburn1998, Einkemmer2013} to discretize spatial derivatives.
The dG methods have the advantage of being high order accurate and parallelizable.
The nonlinear generalized Poisson equation \reff{eq:QN_slab} translates into a 
symmetric algebraic equation\cite{Yadav2013}, which we solve via a conjugate gradient method.
The resulting algorithm is very well suited for current parallel hardware architectures. 
Our GPU implementation thus allows to solve the nonlinear polarization equation efficiently.
 In time we use an explicit Adams-Bashforth multistep method of $3$rd order. 

We carefully verified our global code with the help of the conservation 
equations \reff{eq:Mass_balance} and \reff{eq:Energy_balance}. In addition, we made
quantitative convergence tests in the $L_2$-norm of density and potential. 
With $300^2$ grid cells, using third order polynomials in each cell, we
ensured that convergence is very well reached in our global simulations.
Note that third order polynomials 
are defined by $4$ coefficients, which makes a total of $(4\cdot 300)^2 = 1200^2$ discretization points.

For the local model \reff{eq:GF_local}  we use a pseudospectral scheme\cite{karniadakis} combined with a $2$nd order 
discretization for the Poisson brackets \cite{arakawa}. 
The diffusive part is integrated implicitly.
The local simulations use $4096^2$ grid points, which also ensures convergence for all 
parameters discussed.

We scanned the parameter space varying $\tau=T_i/T_e$, the initial blob width $\sigma$ and the initial amplitude $\Delta n$.
When comparing global to local simulations, we use equal 
physical parameters and initial conditions.
The major radius is set to $R = 4000\rho_s$ with $\rho_s = \frac{\sqrt{m_iT_e}}{eB_0}$.
We fix the ratio of the effective gravity to the dissipative forces $\frac{(1+\tau)\sigma^3\kappa \Delta n}{\nu^2} = 2\cdot 10^5$ and thereby determine the diffusion coefficient $\nu$ given blob width and amplitude. 
 Note that we also tried to fix the diffusive coefficient to $\nu=10^{-2}\Omega_0\rho_s^2$ and found
 only marginal differences compared to the results presented here. This
 means that we are well in the high Reynolds number regime.
The initial blob position is $x_0=0.25L$, $y_0=0.5L$, and we simulate 
from $0$ to $T_\text{max} = 30\gamma_\text{local}^{-1}$ (both local and global simulations) using 
approximately $30000$ timesteps.
Unless otherwise indicated, we fix these parameters throughout the rest of this paper. 

\subsection{Cold ion limit} \label{sec:zlr}
First, we present results from simulations with $\tau=0$. The gyroaveraging operators reduce to $ \Gamma_1 =1$ and $\frac{1}{\tau}(\Gamma_0-1) = \rho_s^2\np^2$, respectively, and hence FLR effects are absent from the models.
In this limit the global model Eqs.~\reff{eq:GF_slab} is a superset of
the local model Eqs.~\reff{eq:GF_local}. Therefore, the global model can be used to test the validity
of the local model in this limit. For small amplitudes we expect the global and local models to show similar results. 
In fact we can use the limit $\frac{\Delta n}{n_0}\ll 1$ as a consistency check for our numerical implementations. 

\begin{figure}[htpb]
    \centering
    \includegraphics[width= 0.95\textwidth]{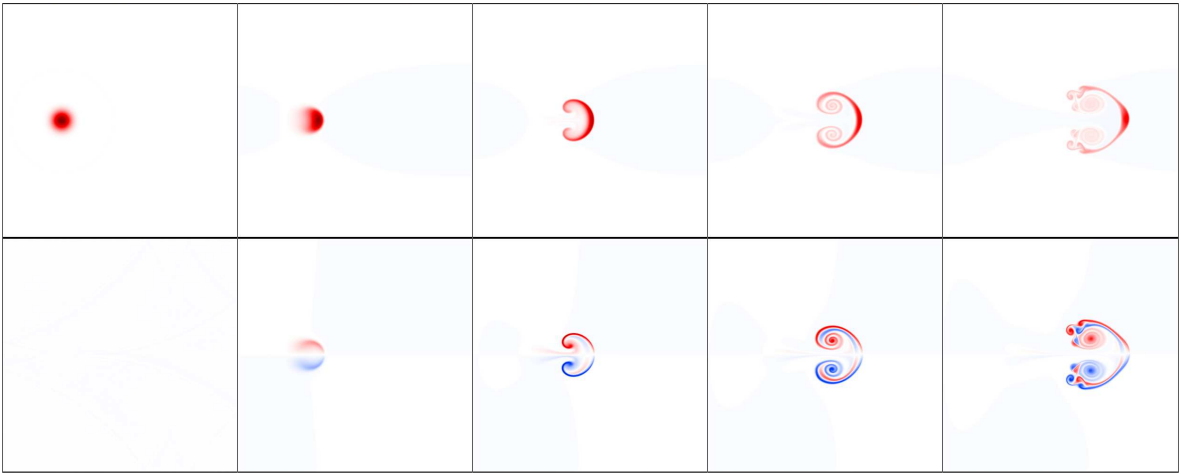}
    \caption{ Density $n$ (top) and vorticity $\np^2 \phi/B_0$ (bottom) of global blob for $\tau=0$, $\sigma=10\rho_s$, and $\Delta n=4n_0$.
        The first column corresponds to $t=0$. Going from left to right, the time increment is $500\Omega_0^{-1}$. The color scales remains constant.
     }
    \label{fig:t0s10a040_global}
\end{figure}

We first raise the question whether the nonlinearity qualitatively changes the 
blob evolution into a mushroom like structure, which was observed previously in local models\cite{Garcia_Bian_Fundamensky_POP_2006}.
Fig. \ref{fig:t0s10a040_global} shows a global simulation with initial blob width $\sigma = 10\rho_s$ and amplitude $\Delta n=4n_0$.
What is shown are contour plots of the particle density and the magnetic field-aligned component of the \ExB-vorticity $\bhat \cdot \nabla\times \vec u_E \approx \np^2\phi/B_0$. Here and in following plots we 
always show the total simulation domain of $(40\sigma)^2$. In the initial phase of the evolution the interchange 
drive term creates a vorticity dipole that  
accelerates the blob radially. 
The dipole accelerates the blob center faster 
in the radial direction than the blob front and its edges. This then leads to a steepening and vertical stretching of 
the blob front. The resulting short length scales are subject to strong diffusion, which in turn leads to a decay of the maximum amplitude. The ultimate result is the 
characteristic mushroom shape with a fast moving blob cap and two
lobes that roll-up and are subject to turbulent mixing. A thorough discussion of these
phenomena is given in ref. \cite{Garcia_Bian_Fundamensky_POP_2006}.

We observe that all our global simulations for zero ion temperature retain this behaviour, in particular the up-down symmetry as seen in Fig. \ref{fig:t0s10a040_global}.
The reason is that the nonlinearity $\nabla N\cdot \nabla\phi$ in the polarization equation \reff{eq:QN_slab}
is small since gradients in $N$ and $\phi$ are mostly perpendicular. 
Note that both the local, as well as the global model contain the symmetry braking
term $\kappa\partial_y n$. This is seen by considering the symmetries in the equations \reff{eq:Elec_Cont_slab} with \reff{eq:vorticity} and \reff{eq:Elec_Cont_local} together with \reff{eq:VortLWL_DF_local} respectively. This term is however small as long as $\rho_s\sqrt{\frac{\kappa}{\sigma}} \ll 1$.

\begin{figure}[htpb]
    \centering
    \includegraphics[width= 0.9\textwidth]{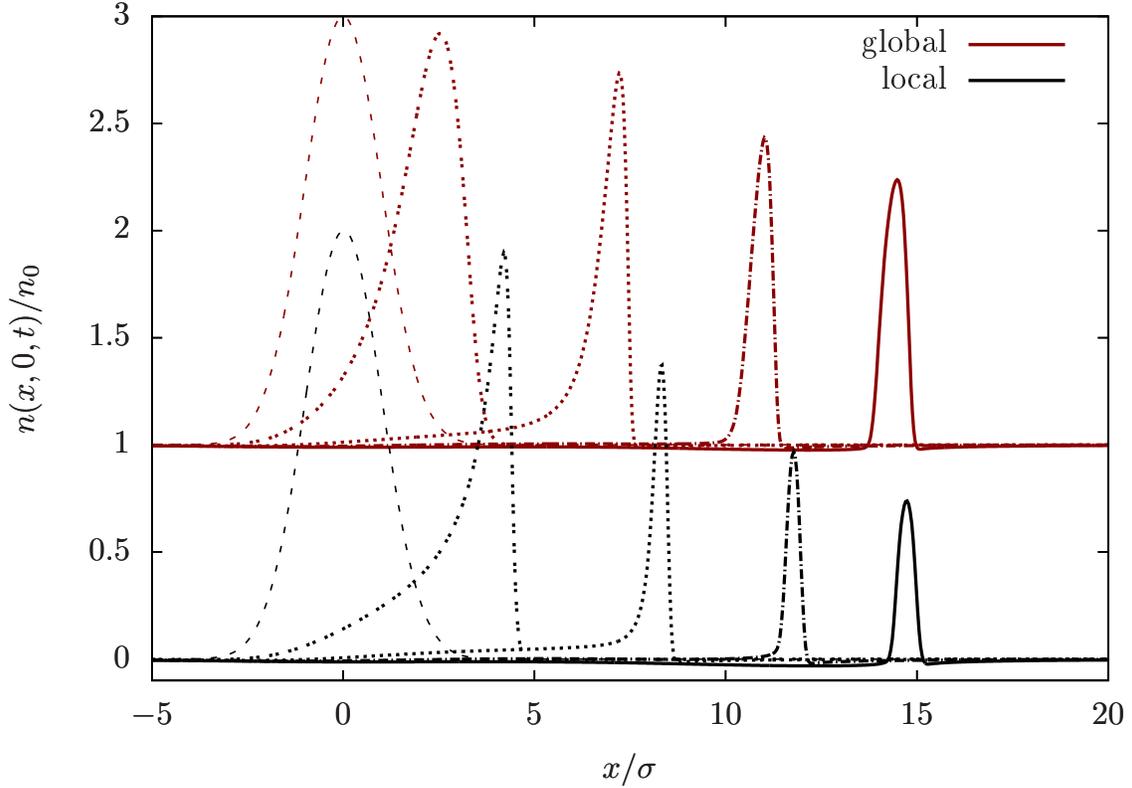}
    \caption{ Radial particle density profiles of local and global blob at $y=0$ for $\tau=0$, $\sigma=10\rho_s$, and $\Delta n=2n_0$ at various timesteps.
        The first dashed line shows the initial blob. Going from left to right, the time increment is $500\Omega_0^{-1}$.
     }
    \label{fig:t0s10a020_radial}
\end{figure}
In order to determine if and what ``global effects'' are present in our simulations, we 
need to present more quantitative results. 
We show radial profiles taken at the symmetry axis $y=0$ in Fig. \ref{fig:t0s10a020_radial}, where we 
compare a global high amplitude simulation to 
a simulation of the local model with equal parameters. Note that we reset the origin of 
the coordinate system to the initial blob position. We observe that the global blob is actually much slower than the local blob
in the initial phase of the evolution. We also observe a weaker radial density gradient
at the global blob front when compared to the very steep local one. 
This results in a reduced particle density diffusion for the global blob. 
While the global blob keeps a high maximal amplitude at later times, the local blob quickly looses more than half of its initial amplitude and slows down.
Both blobs thus travel almost the same distance after $2000\Omega_0^{-1}$, yet at this point in time the
amplitude of the global blob is twice as high as the local one. We conclude that 
the global model must indeed be used to simulate blob convection in this regime. 
\begin{figure}[htpb]
    \begin{center}
        \subfloat[maximum amplitude]{
        \includegraphics[width= 0.5\textwidth]{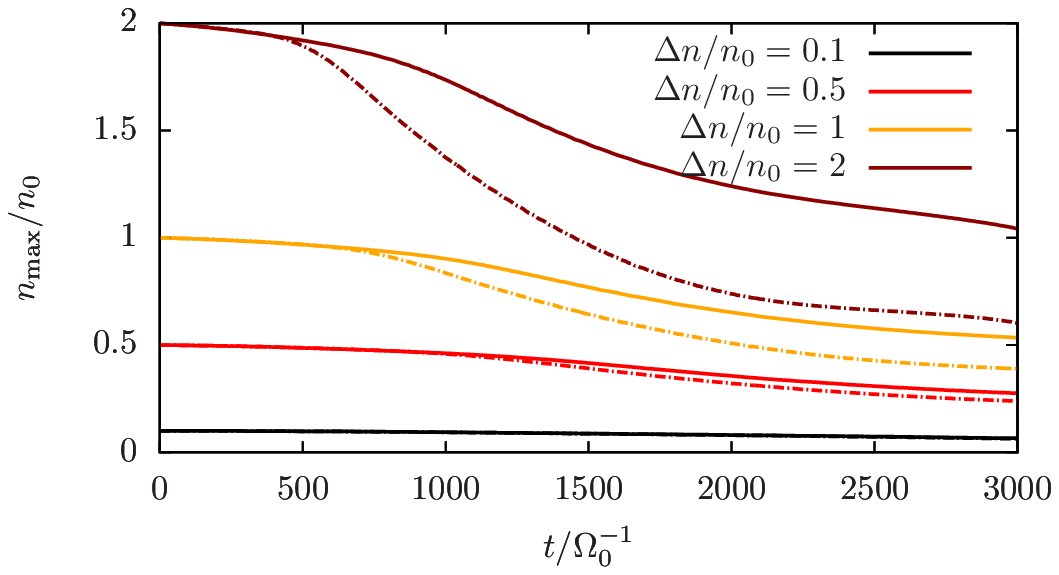}
        \label{fig:t0s10_ampa}
    }
    \subfloat[radial maximum amplitude position]{
        \includegraphics[width= 0.5\textwidth]{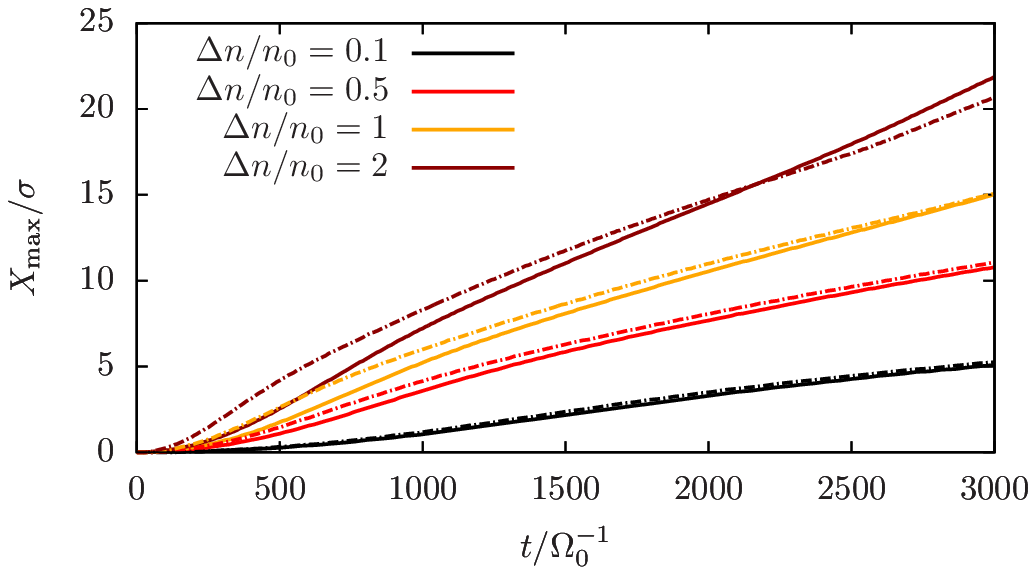}
        \label{fig:t0s10_ampb}
        }
    \end{center}
    \caption{ Maximum amplitude (a) and radial maximum amplitude position (b) for $\tau=0$, and $\sigma=10\rho_s$, and various initial amplitudes as a function of time. Solid lines show global, broken lines local simulations. 
     }
    \label{fig:t0s10_amp}
\end{figure}

To quantify our findings further, we plot the maximum amplitude and the radial maximum amplitude position 
for various initial amplitudes in Fig. \ref{fig:t0s10_ampa} and \ref{fig:t0s10_ampb} respectively.
The maximum amplitude at time $t$ is $n_\text{max}(t) := \max_{\vec x\in D}\{n(\vec x,t)-n_0\}$, and $\vec x_{\text{max}}$ denotes the corresponding position.
The curves for low amplitudes almost fall on top of each other as expected. 
We observe that in both the local and the global model the amplitude is reduced with time for all initial amplitudes. 
However, the amplitude in the local model is clearly smaller when compared to the global one, especially for higher initial amplitudes. 
We can also confirm that in the initial phase the radial maximum amplitude positions for global blobs lag behind those of local blobs. Only at later times global blobs catch up and the maximum amplitude positions coincide. 

\begin{figure}[htpb]
    \begin{center}
        \subfloat[Gyro-Bohm scaling]{
            \label{}
            \includegraphics[width= 0.5\textwidth]{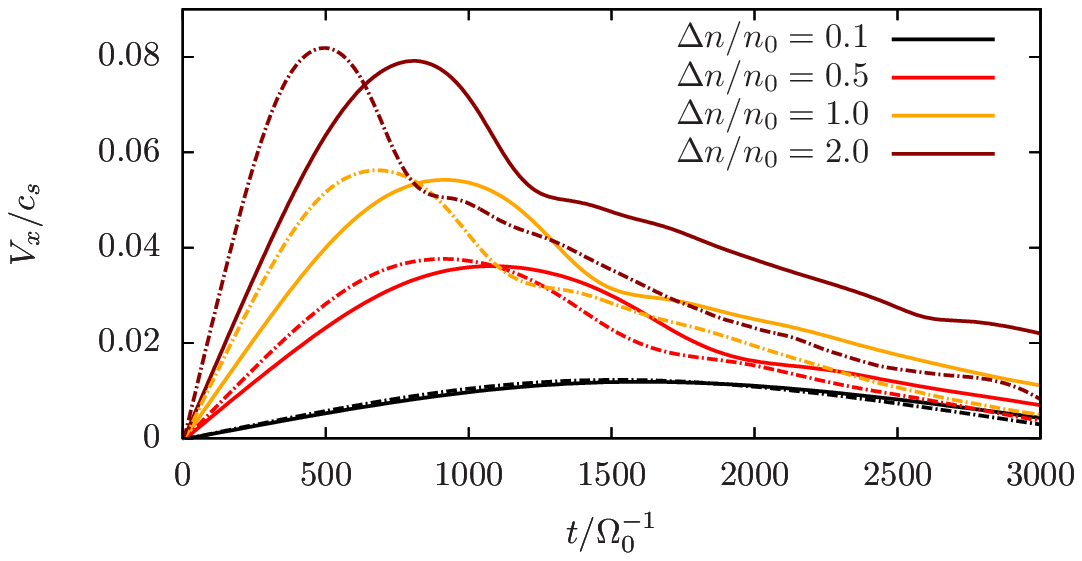}
            \label{fig:t0s10_global_vxa}
        }
        \subfloat[global scaling]{
            \label{}
            \includegraphics[width= 0.5\textwidth]{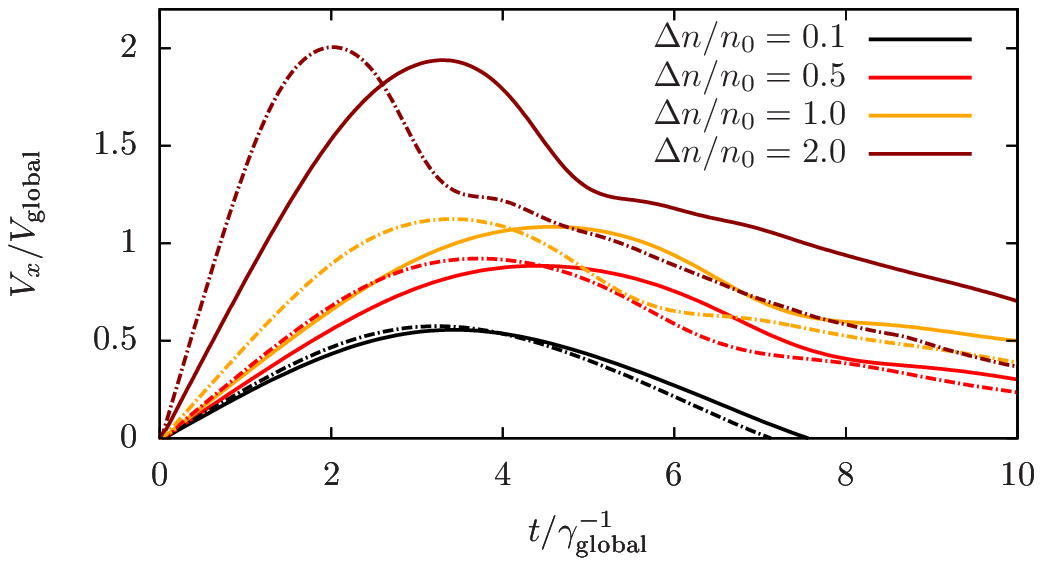}
            \label{fig:t0s10_global_vxb}
        }
    \end{center}
    \caption{ Global and local blob simulations for $\tau=0$ and $\sigma=10\rho_s$. 
        We show the radial center of mass velocity as a function of time normalized by (a) the ion gyration time $\Omega_0^{-1}$ and (b) the ideal global interchange time $\gamma_\text{global}^{-1}$. Solid lines show global, broken lines local simulations, respectively.
     }
    \label{fig:t0s10_global_vx}
\end{figure}

The next step in our discussion is to investigate center of mass positions and velocities.
We define the center of mass of a blob by 
\begin{align}
    \vec X_C := \frac{1}{\int \left[ n -n_0 \right] \dV }\int \vec x \left[ n -n_0 \right] \dV.
    \label{eq:com}
\end{align}
The center of mass velocity, which is also a measure for the advective \ExB-flux\cite{jmad2011FLRBlob}, then follows as
\begin{align}
    \vec V_C := \frac{d}{dt}\vec X_C .
    \label{eq:com_vel}
\end{align}
We plot center of mass velocities of local and global blobs for various amplitudes and fixed blob width $\sigma=10\rho_s$ in Fig. \ref{fig:t0s10_global_vx}. 
We used the standard Gyro-Bohm scaling in Fig. \ref{fig:t0s10_global_vxa}. 
Again, the center of mass velocities for the blobs with the smallest amplitudes almost coincide as expected. 
In accordance with the radial profiles shown in Fig. \ref{fig:t0s10a020_radial} we observe that in the beginning of the blob evolution 
the high amplitude global blobs accelerate less and thus have lower velocities when compared
with local blobs having identical parameters. The local blobs reach their maximum velocity earlier in their
evolution and then quickly decelerate. The global blobs take longer times to reach their 
maximal velocities and retain increased speeds in the later phases. 
This is in line with the global model using the correct ion inertia, while
the local model uses a constant background one. 
However, the maximum velocity is slightly reduced for global, high amplitude blobs.
In order to test whether blob amplitude variations are captured by
the previously derived scaling law for global blob velocities \reff{eq:velocity_scaling_global}, we show the same simulation results using 
the global interchange rate and velocity as scaling parameters in Fig. \ref{fig:t0s10_global_vxb}. The curves do not
fall on top of each other as we might have expected, yet the global scaling seems to capture the 
dynamics fairly well. 

\begin{figure}[htpb]
    \centering
    \includegraphics[width= 0.9\textwidth]{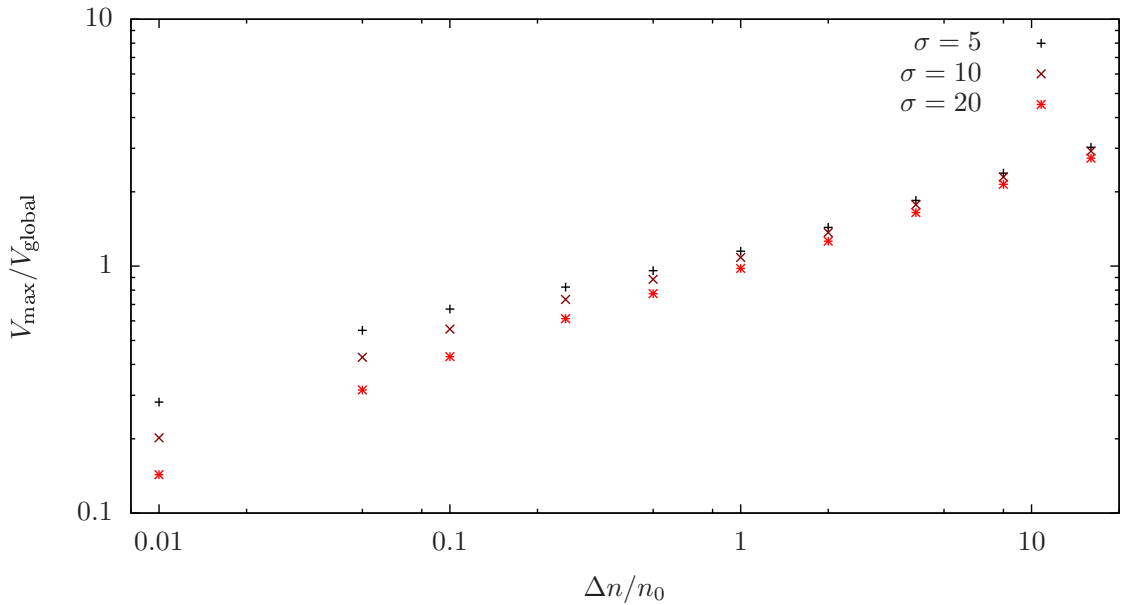}
    \caption{ 
    Global blob simulations for $\tau=0$ and various blob widths. 
    We show the maximum radial velocity scaled by the global interchange velocity \reff{eq:velocity_scaling_global} as a function of amplitude. 
     }
    \label{fig:t0_global_vmax}
\end{figure}

Kube et al. \cite{Kube2011} have used a drift fluid model to describe the behaviour of global blobs in the zero ion temperature limit. The local velocity scaling Eq.~\reff{eq:velocity_scaling_local}
was validated very well for small amplitudes.
We note that their model resembles our model if the term $\nabla \ln N\cdot \nabla \phi$ in the polarization equation is neglected and if $\tau = 0$.
We plot the maximum velocity scaled by the global interchange velocity \reff{eq:velocity_scaling_global} as a function of amplitude in Fig. \ref{fig:t0_global_vmax}. 
The scaling is apparently flawed as neither for low nor for high amplitudes the curves are constant horizontal lines.
Note that \cite{Kube2011} also failed to recover the velocity scaling in the high
amplitude regime although even higher amplitudes than ours were used in the simulations. 
One reason might be that the amplitude of the blobs can be significantly decreased
by the time the maximum velocity is actually reached (cf. Fig. \ref{fig:t0s10_ampa}). The initial amplitude
might thus not be the one that should be used for the plot.
The variation in width is well captured for amplitudes higher than $\Delta n=1n_0$. We remark that \cite{Kube2011} did not
vary the blob width, which was absorbed in their scaling. 
All in all, we see that the amplitude dependence of the velocity scaling in Eq.~\reff{eq:velocity_scaling_global} is not well described by the theoretical estimate.

\subsection{Finite ion temperature} \label{sec:flr}
We now discuss simulations taking a constant finite ion temperature into account. 
Local simulations with amplitude $\Delta n=0.5n_0$ including FLR effects were first published in \cite{jmad2011FLRBlob}. It was found that the blob dynamics is significantly
altered by retaining FLR effects in the model. Blobs move radially as well as poloidally and stay more coherent compared to zero ion temperature simulations.

Our main point in this section is to investigate differences between the local and the global gyrofluid model. 
As described in the theory section \ref{sec:comparison}, FLR 
corrections to the polarization density are only present in the local model. 
These corrections enter as powers of $(\rho_i k_\perp)^2$ as seen e.g. in Eq. \reff{eq:N_LWL_local}.  
However, only the global model retains the nonlinear polarization density
in the polarization equation. 

\begin{figure}[htpb]
    \centering
    \includegraphics[width= 0.95\textwidth]{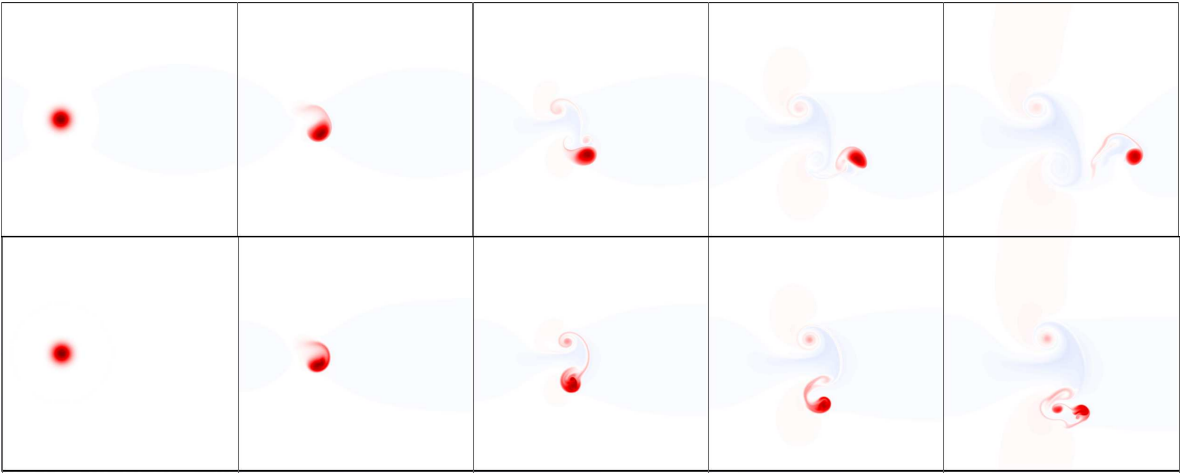}
    \caption{ Particle density $n$ of local (top) and global (bottom) blob for $\tau=4$, $\sigma=5\rho_s$, $\Delta n=0.5n_0$. 
    The first column corresponds to $t=0$. Going from left to right, the time increment is $475\Omega_0^{-1}$. The color scale remains constant.
     }
    \label{fig:t4s05a005_density}
\end{figure}

\begin{figure}[htpb]
    \centering
    \includegraphics[width= 0.95\textwidth]{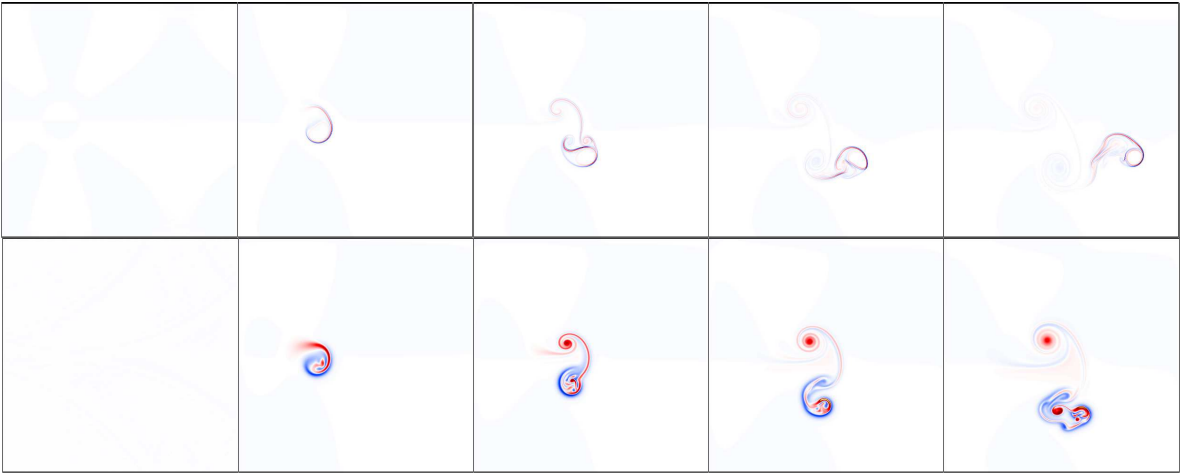}
    \caption{ Vorticity $\np^2\phi/B_0$ of local (top) and global (bottom) blob for $\tau=4$, $\sigma=5\rho_s$, $\Delta n=0.5n_0$. 
    Time increment is $475\Omega_0^{-1}$. Note that the color scale for the global vorticity is $20$ times lower than that of the local one.
     }
    \label{fig:t4s05a005_vorticity}
\end{figure}

As a first example we choose $\tau = 4$, $\sigma = 5\rho_s$, and $\Delta n=0.5n_0$. From 
both local and global simulations we plot the particle density and vorticity fields in Fig. \ref{fig:t4s05a005_density} and \ref{fig:t4s05a005_vorticity} respectively. 
We loosely estimate $(\rho_i k_\perp)^2 = \left(\frac{\rho_i \Delta n}{\sigma(n_0+\Delta n)}\right)^2 \approx 0.02 \ll 1 $ and thus expect only weak FLR effects, at least during
the first timesteps. 
From the particle density plots we see that the qualitative blob movement in the initial phase is indeed similar in both cases. 
Both blobs accelerate radially as well as in the poloidal direction, which in our case
is in fact the $\hat b \times \nabla B$ direction, where $\hat b$ points out of the paper (cf. also \cite{jmad2011FLRBlob}).
However, in the later phase of the evolution clear differences can be seen. 
The global blob is slower and looses more mass to 
dissolving vortices that separate from the main blob. The local
blob travels much farther in the radial direction and retains its initial form during the whole simulation period. Also the 
poloidal movements differ. The local blob reverses its poloidal velocity twice,
the global blob only once. 

In Fig. \ref{fig:t4s05a005_vorticity} we observe
very pronounced differences in the vorticity between the local and the global model. The local blob quickly develops a strong
and highly localized sheared flow around the blob. Note that the color scale for the local 
case is $20$ times higher than that for the global case. This 
sheared flow is the reason for the enhanced stability of the blob shape, which is persistent over
the whole simulation period\cite{jmad2011FLRBlob}. 
The global blob lacks such a violent vorticity roll-up and is thus unable 
to maintain its shape loosing mass in Kelvin-Helmholtz like vortices
at later times. Moreover, we observe more internal structures in the vorticity field.

A possible explanation for the observed differences between the local and global 
vorticity fields could be the absence of the $\nabla N\cdot \nabla_\perp\phi$ nonlinearity
in the local polarization equation (\ref{eq:QN_slab}). 
We observe that the particle density and the electric potential gradients align
at the blob edge. 
However, a closer inspection reveals that the particle density amplitude is very small where the gradients align, so the effects of the nonlinearity is expected to be small. 
Another possible explanation is the absence of FLR corrections to the polarization 
density in the global polarization equation. 
These enter the local polarization equation as:
\begin{align*}
\Gamma_0-1 = (\rho_i\np)^2\left[ 1 + (\rho_i\np)^2 + \dots \right].
    \label{}
\end{align*}
To check whether the differences in the vorticity fields
are indeed due to this factor, we repeated our local simulations replacing $\Gamma_0-1$ by a Laplacian in equation \reff{eq:QN_local}:
\begin{align}
    \Gamma_1 \tilde N + \frac{e n_0}{T_e} \rho_s^2\np^2\phi = \tilde n .
\label{eq:modified}
\end{align}
We denote this as the modified local model. 
\begin{figure}[htpb]
    \centering
    \includegraphics[width= 0.9\textwidth]{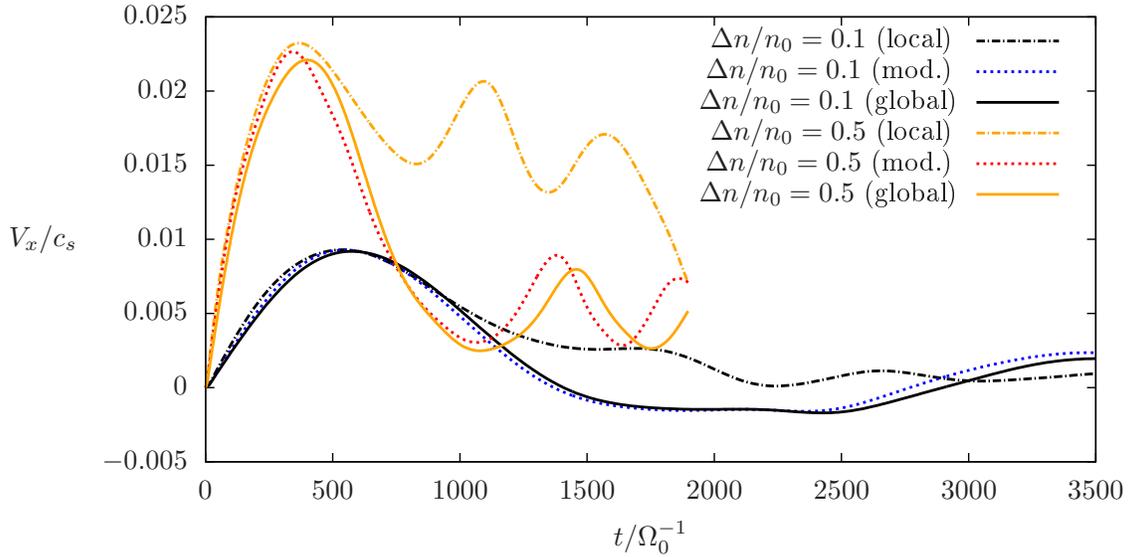}
    \caption{ Comparison of global and local blobs for $\tau=4$ and $\sigma=5\rho_s$. 
    In addition, we modified the local model replacing $\Gamma_0-1$ by $\rho_i^2\np^2$ in the 
    polarization equation (cf. Eq. \reff{eq:modified}).
    We show center of mass velocity as a function of time. 
     }
    \label{fig:t4s05_vx}
\end{figure}
We plot the center of mass velocities of local, modified, and global blobs in Fig. \ref{fig:t4s05_vx}. 
As in the 
zero ion temperature case the velocity in the initial phase is slightly higher in both local models than in the global model.
At later times we see that
the local blob is up to two times faster than its global and modified counterparts. 
As a side remark we note that velocity peaks coincide with poloidal turns. 
The global blob as well as the modified local blob quickly slows 
down after the first velocity peak, probably because the surrounding velocity field, which
prevents blob fragmentation,
is not as strong in the global and modified blob as it is in the local blob (cf. Fig. \ref{fig:t4s05a005_vorticity}). From Fig. \ref{fig:t4s05_vx} we conclude that  
the FLR corrections to the polarization density are indeed responsible for the
different behaviour of local and global blobs in the late phase of the blob evolution. 
All in all, we conclude that for low amplitudes, small blob widths, and high 
ion temperatures, the local model is the preferable
model since FLR corrections are consistently maintained in the polarization equation,
which is not the case in the global model. 

\subsection{High amplitude blobs} \label{sec:high}
\begin{figure}[htpb]
    \centering
    \includegraphics[width= 0.95\textwidth]{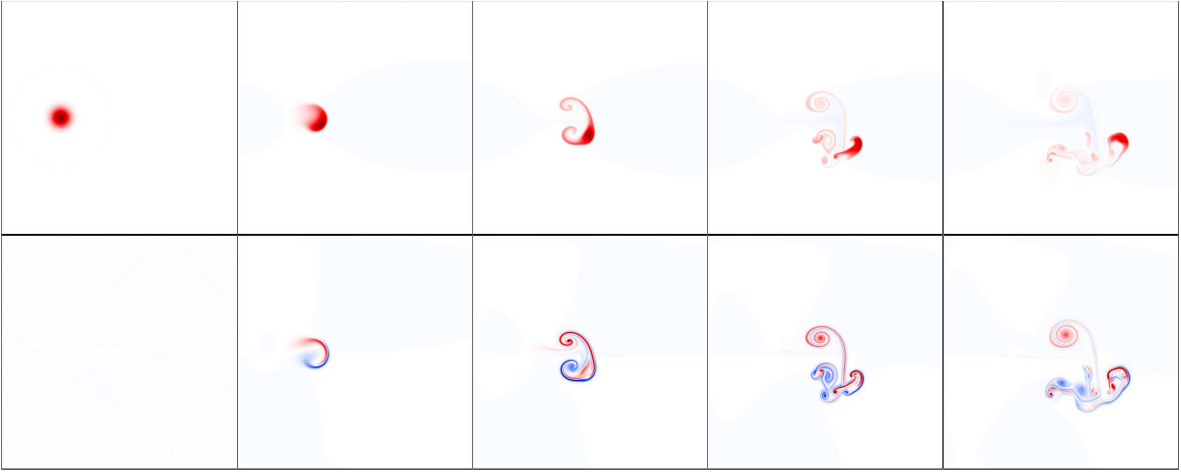}
    \caption{ 
Density $n$ (top) and vorticity $\np^2\phi/B_0$ (bottom) plot of global blob for $\tau=2$, $\sigma=10\rho_s$, and $\Delta n=2n_0$.
The first column corresponds to $t=0$. Going from left to right, the time increment is $430\Omega_0^{-1}$. The color scales remains constant. 
     }
    \label{fig:t2s10a020_global}
\end{figure}

We now show global, high amplitude blob simulations with moderate FLR effects. 
In this parameter regime the local model is not valid.
We reduce the ion temperature and increase the blob width compared to the previous section. 
This reduces the ratio of ion gyroradius to gradient
length scale, which measures the strength of FLR effects as discussed in the previous section.
We exemplarily show contour plots of the particle density and vorticity for $\tau=2$, $\sigma=10\rho_s$, and 
$\Delta n=2n_0$ in Fig. \ref{fig:t2s10a020_global}.
The evolution is best described as a mixture of the high temperature blobs in the last section and the cold ion blobs in section \ref{sec:zlr}. The blob accelerates radially as well as poloidally
in the initial phase with the vorticity slightly rolling up. Two side-arms with a 
pronounced cap develop afterwards, which resembles the mushroom shapes of cold ion blobs. 
In the poloidal turn the blob becomes stretched and separates from its lobes, streaming
upwards thereafter. 
Scanning the parameter range we found that  
the blob evolution either becomes more mushroom like for low ion temperature and large blob widths or more compact for high ion temperature and small widths.
Yet, before we come back to this observation of blob shapes, we want to examine radial profiles, maximum
amplitude position, and center of mass velocities as we did in section \ref{sec:zlr}.
\begin{figure}[htpb]
    \centering
    \includegraphics[width= 0.9\textwidth]{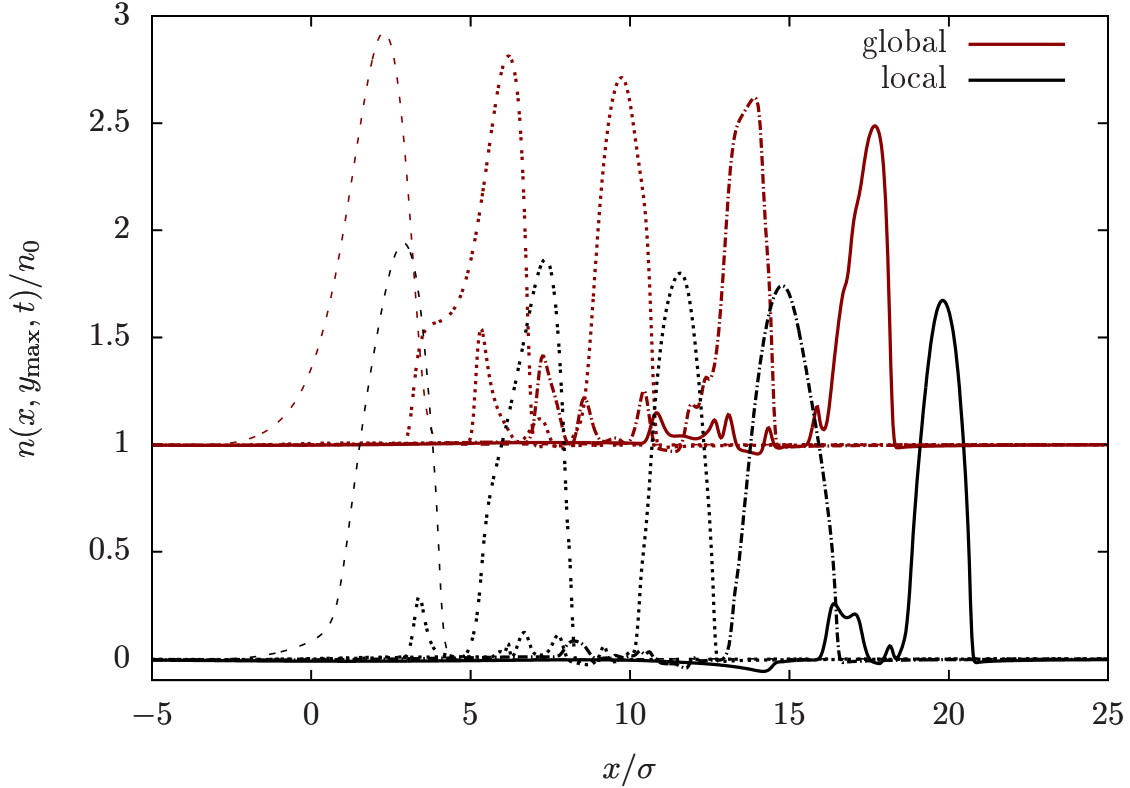}
    \caption{ Radial particle density profiles for $\sigma=10\rho_s$, $\Delta n=2n_0$, and $\tau=2$. The profiles are taken at the poloidal maximum amplitude position at time (from left to right) $287$, $2\cdot287$, $3\cdot287$, $4\cdot287$,  and $5\cdot287\Omega_0^{-1}$. 
     }
    \label{fig:t2s10a020_radial}
\end{figure}

First, we show radial profiles of the plasma density in Fig. \ref{fig:t2s10a020_radial}.
Since the up-down symmetry of the cold ion blobs is broken, we take the profiles
at the poloidal maximum amplitude position of the blob. 
Profiles from local and global models resemble each other. 
In the vicinity of the maximal particle density the profiles are approximately Gaussian shaped with a fluctuating, low amplitude tail.
There are slightly more fluctuations present
in the global curves. 
When compared to the profiles in Fig. \ref{fig:t0s10a020_radial}, where $\tau=0$, we see that 
the low temperature blobs have steeper profiles than the blobs with $\tau = 2$. 
Also the loss of maximum amplitude is not as pronounced for 
the warm ion case as it is for the cold ion case. Furthermore, the local blob always stays ahead of the global one. 

\begin{figure}[htpb]
    \centering
    \includegraphics[width= 0.9\textwidth]{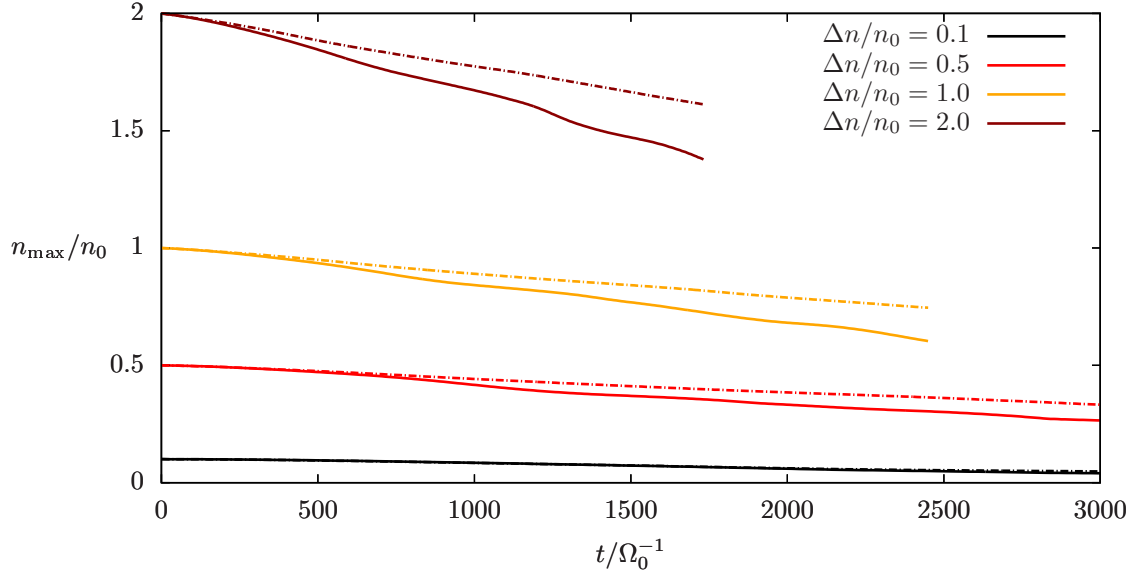}
    \caption{ Maximum amplitude for $\sigma=10\rho_s$ and $\tau=2$ as a function of time.
    Solid lines show global, broken lines local simulations.  
     }
    \label{fig:t2s10_max_amp}
\end{figure}

Next, we plot the maximum amplitude as a function of time in Fig. \ref{fig:t2s10_max_amp}. 
As expected the small amplitude curves coincide. 
Contrary to Fig. \ref{fig:t0s10_ampa} in section \ref{sec:zlr}, which is
the zero ion temperature version of Fig. \ref{fig:t2s10_max_amp}, we find
that now local blobs retain their amplitude better than their global counterparts. 
With regard to the preceding discussion of blob stability this does not come as 
a surprise. Local blobs stay coherent during the whole simulation time and 
keep mass and hence amplitude almost constant. 

In order to test the global velocity scaling \reff{eq:velocity_scaling_global}, we examine the radial center of mass
velocity as a function of time. 
\begin{figure}[htpb]     
    \begin{center}
        \subfloat[ $\Delta n/n_0=2$]{
            \includegraphics[width= 0.5\textwidth]{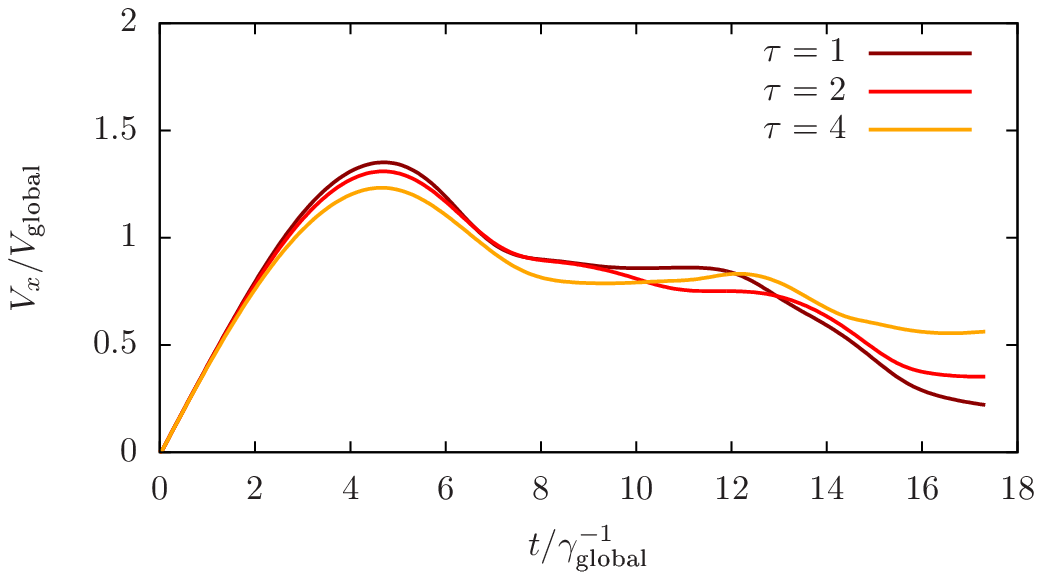}
            \label{fig:t2s10a}
        }
        \subfloat[$\tau=2$]{
            \includegraphics[width= 0.5\textwidth]{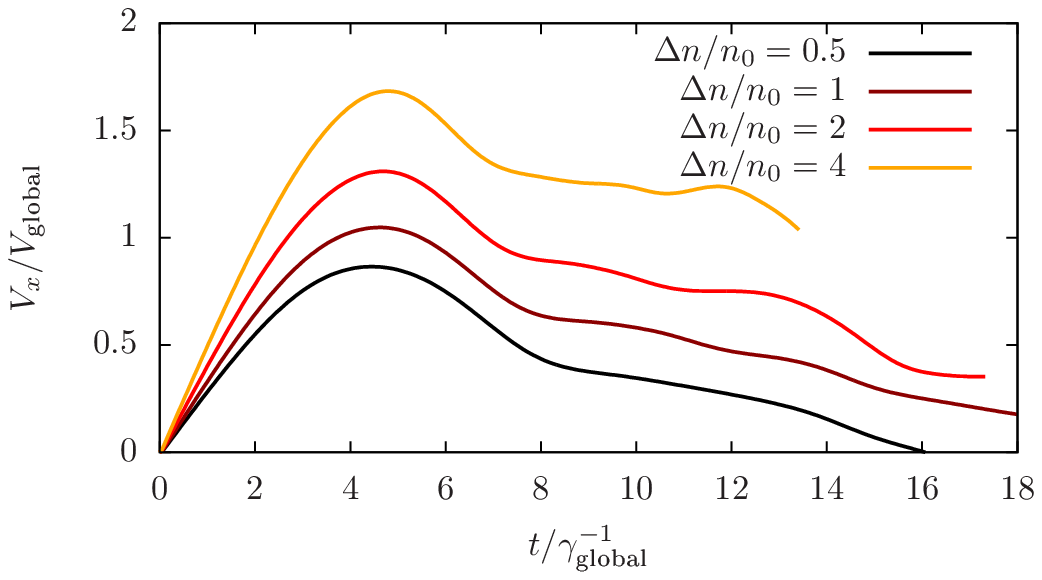}
            \label{fig:t2s10b}
        }
    \end{center}
    \caption{ Radial center of mass
    velocity as a function of time, for $\sigma=10\rho_s$. We vary the ion temperature 
    for fixed amplitude $\Delta n/n_0 = 2$ (a) and the amplitude for fixed ion temperature $\tau = 2$ (b). 
     }
     \label{fig:t2s10}
\end{figure}
In Fig. \ref{fig:t2s10a} we see that the global scaling captures the 
ion temperature variation very well. The variation of amplitude is, like in section \ref{sec:zlr}, only partly captured. In both figures we see that the velocity in the
initial phase increases almost linearly until it reaches a maximum and decreases
again. 
At about $7\gamma_{\text{global}}^{-1}$ there is a sudden transition where the blob velocity stabilizes at 
an almost constant value until it finally drops down to smaller values again.
When 
inspecting the particle density plots in Fig. \ref{fig:t2s10a020_global}, the transition takes place at the point where the lobes of the blob start to curl and roll up. The
second drop of velocity occurs when the blob starts to fragment at about $13\gamma_\text{global}^{-1}$.

We now come back to the observation that blobs have a tendency to either develop 
a mushroom shape, to retain a more coherent blob-like structure, or a mixture of both.
We use the definition of blob compactness\cite{jmad2011FLRBlob}
\begin{align}
    I_C(t) := \frac{\int_D \dV (n(x,y,t)-n_0)h(x,y,t)}{\int_D\dV(n(x,y,0)-n_0)h(x,y,0)},
    \label{eq:compactness}
\end{align}
where $h$ is defined as a Heaviside function 
\begin{align}
    h(x,y,t):=\begin{cases}
        1\quad\text{     if }(x-x_\text{max}(t))^2+ (y-y_\text{max}(t))^2 < \sigma^2, \\
        0\quad\text{     else.}
    \end{cases}
    \label{eq:heaviside}
\end{align}
The integration is thus performed on a circular field of radius $\sigma$ around the maximum amplitude position.

$I_C$ is a measure for the ability of the blob to retain its form and mass. A small
compactness means that the blob has lost most of its initial mass or is
spread out over a large area. The mushroom shapes in section \ref{sec:zlr} should e.g. have a small compactness.  
A high compactness means that the blob preserves its initial particle density.
The high ion temperature
blobs in section \ref{sec:flr} should correspondingly have a high compactness.
\begin{figure}[htpb]
    \centering
    \includegraphics[width= 0.9\textwidth]{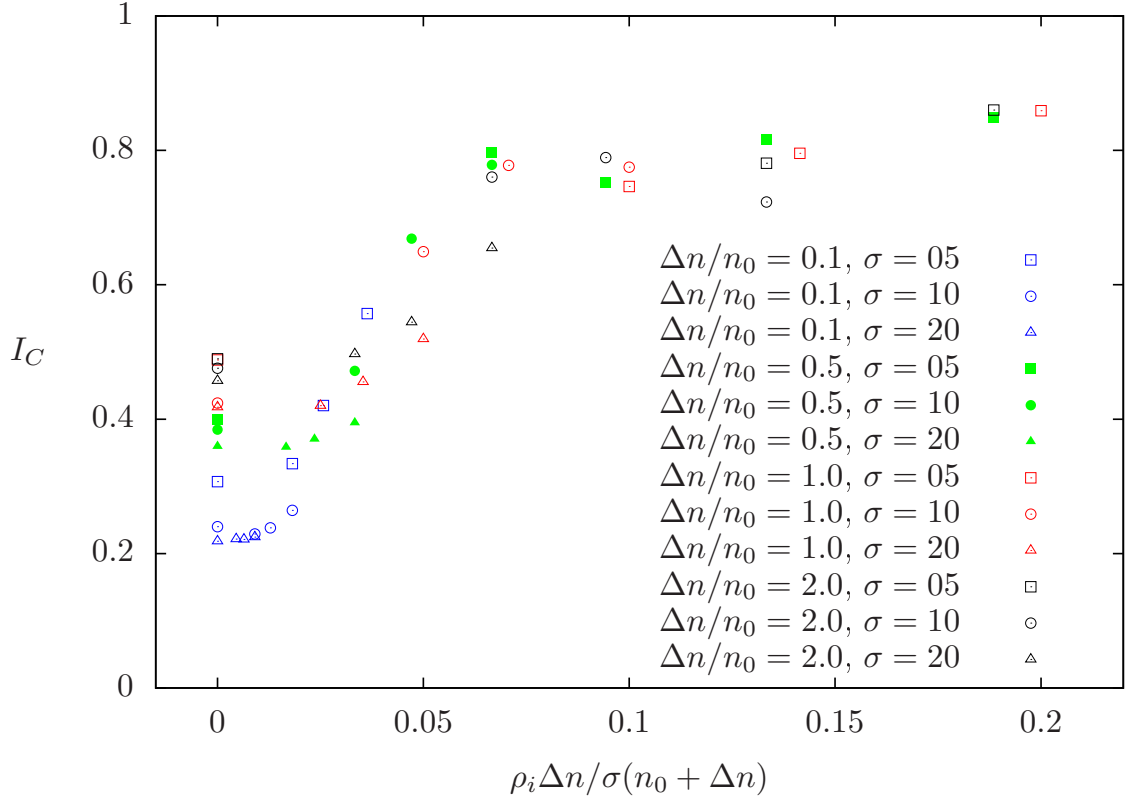}
   \caption{ Blob compactness $I_C$ of global blobs as a function of FLR strength at time
   $t=10\gamma_\text{global}^{-1}$ for various amplitudes and blob widths.
     }
    \label{fig:figure17}
\end{figure}
In Fig. \ref{fig:figure17} we show the blob compactness at time $t=10\gamma_\text{global}^{-1}$ as a function of the
FLR strength modeled by the control parameter
\begin{align}
   r = \frac{\rho_i}{\sigma}\frac{ \Delta n}{(n_0 + \Delta n)}.
    \label{eq:factorr}
\end{align}
$r$ is the ratio between the ion gyroradius and the initial gradient length scale, which
we have already used in the preceding discussions.
 In line with 
the results presented in \cite{jmad2011FLRBlob} we identify a transition between 
$r=0$  and $r=0.075$ where $I_C$ increases significantly. For higher values of $r$ the 
compactness constantly fluctuates around $0.8$ for all parameters
investigated in this regime. For low values of $r$ the compactness
is a factor $2-3$ times smaller, showing that blob mass in this regime will rather spread out or diffuse away. Furthermore, blobs with very low FLR effects show 
a significant variation of compactness when amplitude is varied. The smallest
values for $I_C$ in our plot can be observed for the low amplitude $\Delta n=0.1n_0$. When amplitude is
increased, the blob compactness increases as well. 

We remark that the cold ion simulations in section \ref{sec:zlr} are found on the left
side of the plot at $r=0$. The high temperature simulations in \ref{sec:flr} are on 
the far right side, while the simulations presented in this section are found
in between. 
Our plot thus shows that $r$, being a combination of blob parameters $\tau$, $\Delta n$, and $\sigma$ only, is a very good indicator of whether a blob can retain its mass during its evolution 
or not.

\section{Conclusion} \label{sec:conclusion}
We showed that we can numerically solve the nonlinear polarization equation
in the context of a mass and energy conserving, 2D gyrofluid model. 
The model was used to investigate blob dynamics of seeded blobs in the tokamak
scrape-off-layer.
We identified two regimes of blob convection. Blobs, defined as the vicinity of the maximal amplitude position, quickly loose
mass in the first and retain their mass in the second regime as they propagate radially. 
Our simulations indicate that over a wide range of parameters, namely ion temperature, initial blob width, and initial blob amplitude, these two regimes are characterised by the 
ratio of ion gyroradius to the initial gradient scale length.
This ratio is interpreted as a measure for the strength of FLR effects. 
Blobs with a low ratio belong to the first, blobs with strong FLR effects belong to the second regime. 
		
Furthermore, we investigated the importance of using a global, fully nonlinear model
in contrast to a local thin layer approximation for blob simulations.
For low ion temperatures and high blob amplitudes we find that global blobs stay more coherent and have an increased cross-field transport compared 
to local model simulations. 
The amplitude in global simulations remains significantly higher than in local simulations with equal initial amplitudes.
When the ion temperature is comparable to the electron temperature, global blob simulations show a decreased cross-field transport in comparison with local blob simulations.
Yet, for low amplitudes we find that the local model is preferable
since FLR corrections to the polarization density are absent 
from the global model.

\section*{Acknowledgements} 	
We would like to thank Ole Meyer for helpful comments on the manuscript.
This work was supported by the Austrian Science Fund (FWF) W1227-N16 and Y398, and by 
the European Commission under the Contract of Association between EURATOM and \"OAW, carried out within the framework of the European Fusion Development Agreement (EFDA).
The computational results presented have been achieved in part using the Vienna Scientific Cluster (VSC). This work was supported by the Austrian Ministry of Science BMWF as part of the UniInfrastrukturprogramm of the Focal Point Scientific Computing at the University of Innsbruck.

%


\end{document}